\documentclass[a4paper,fleqn,usenatbib]{mnras/mnras}

\usepackage{mathptmx}

\usepackage[T1]{fontenc}
\usepackage{ae,aecompl}

\usepackage{graphicx}	
\usepackage{amsmath}	
\usepackage{amssymb}	
\usepackage[utf8]{inputenc}
\usepackage[flushleft]{threeparttable} 
\usepackage{hyperref} 
\usepackage{lineno}

\usepackage[nodisplayskipstretch]{setspace}
\usepackage{eso-pic}

\AddToShipoutPictureBG*{%
  \AtPageUpperLeft{%
    \hspace{0.75\paperwidth}%
    \raisebox{-3.5\baselineskip}{%
      \makebox[0pt][l]{\textnormal{DES 2016-0177}}
}}}%

\AddToShipoutPictureBG*{%
  \AtPageUpperLeft{%
    \hspace{0.75\paperwidth}%
    \raisebox{-4.5\baselineskip}{%
      \makebox[0pt][l]{\textnormal{Fermilab PUB-16-220}}
}}}%

\title[VDES J2325-5229 a z=2.7 gravitationally lensed quasar]{VDES J2325-5229 a z=2.7 gravitationally lensed quasar discovered using morphology independent supervised machine learning}

\author[F. Ostrovski, R.G. McMahon et al.]
{\parbox{\textwidth} 
{Fernanda Ostrovski$^{1,2,3}$
Richard G. McMahon$^{1,2}$,
Andrew J. Connolly$^{1,4}$,
Cameron A. Lemon$^{1,2}$, 
Matthew W. Auger$^{1}$, 
Manda Banerji$^{1,2}$,
Johnathan M. Hung$^{1,5}$, 
Sergey E. Koposov$^{1}$,
Christopher E. Lidman$^{6,7}$,
Sophie L. Reed$^{1,2}$,
Sahar Allam$^{8}$, Aur{\'e}lien Benoit-L{\'e}vy$^{9,10,11}$, Emmanuel Bertin$^{9,11}$, David Brooks$^{10}$, Elizabeth Buckley-Geer$^{8}$, Aurelio Carnero Rosell$^{12,13}$, Matias Carrasco Kind$^{14,15}$, Jorge Carretero$^{16,17}$, Carlos E. Cunha$^{18}$, Luiz N. da Costa$^{12,13}$, Shantanu Desai$^{19,20}$,
H. Thomas Diehl$^{8}$, J\"{o}rg P. Dietrich$^{19,20}$, August E. Evrard$^{21,22}$, David A. Finley$^{8}$, Brenna Flaugher$^{8}$, Pablo Fosalba$^{16}$, Josh Frieman$^{8,23}$, David W. Gerdes$^{22}$, Daniel A. Goldstein$^{24,25}$, Daniel Gruen$^{18,26}$, Robert A. Gruendl$^{14,15}$, Gaston Gutierrez$^{8}$, Klaus Honscheid$^{27,28}$, David J. James$^{29}$, Kyler Kuehn$^{30}$, Nikolay Kuropatkin$^{8}$, Marcos Lima$^{12,31}$, Huan Lin$^{8}$, Marcio A. G. Maia$^{12,13}$, Jennifer L. Marshall$^{32}$, Paul Martini$^{27,33}$, Peter Melchior$^{34}$, Ramon Miquel$^{17,35}$, Ricardo Ogando$^{12,13}$,  Andr{\'e}s Plazas Malag{\'o}n$^{36}$, Kevin Reil$^{26}$, Kathy Romer$^{37}$, Eusebio Sanchez$^{38}$, Basilio Santiago$^{12,39}$, Vic Scarpine$^{8}$, Ignacio Sevilla-Noarbe$^{38}$, Marcelle Soares-Santos$^{8}$, Flavia Sobreira$^{12,40}$, Eric Suchyta$^{41}$, Gregory Tarle$^{22}$, Daniel Thomas$^{42}$, Douglas L. Tucker$^{8}$, Alistair R. Walker$^{29}$}\\~\\
Affliations at end of paper.
}

\date{Accepted 2016 November 14. Received 2016 November 7; in original form 2016 July 5}

\pubyear{2016}

\begin{document}
\label{firstpage}
\pagerange{\pageref{firstpage}--\pageref{lastpage}}
\maketitle

\begin{abstract}

We present the discovery and preliminary characterization of a gravitationally lensed quasar with a source redshift $z_{s}=2.74$ and image separation of $2.9''$ lensed by a foreground $z_{l}=0.40$ elliptical galaxy. Since optical observations of gravitationally lensed quasars show the lens system as a superposition of multiple point sources and a foreground lensing galaxy, we have developed a morphology independent multi-wavelength approach to the photometric selection of lensed quasar candidates based on Gaussian Mixture Models (GMM) supervised machine learning.  Using this technique and $gi$ multicolour photometric observations from the Dark Energy Survey (DES), near IR $JK$ photometry from the VISTA Hemisphere Survey (VHS) and WISE mid IR photometry, we have identified a candidate system with two catalogue components with $i_{AB}=18.61$ and $i_{AB}=20.44$ comprised of an elliptical galaxy and two blue point sources. Spectroscopic follow-up with NTT and the use of an archival AAT spectrum show that the point sources can be identified as a lensed quasar with an emission line redshift of $z=2.739\pm0.003$ and a foreground early type galaxy with $z=0.400\pm0.002$. We model the system as a single isothermal ellipsoid and find the Einstein radius $\theta_E \sim 1.47''$, enclosed mass $M_{enc} \sim 4 \times 10^{11}$M$_{\odot}$ and a time delay of $\sim$52 days. The relatively wide separation, month scale time delay duration and high redshift make this an ideal system for constraining the expansion rate beyond a redshift of 1.


\end{abstract}

\begin{keywords}
gravitational lensing: strong -- quasars: general -- methods: observational -- methods: statistical
\end{keywords}



\section{Introduction}

The discovery of the first strong gravitational lens \citep{Walsh+1979}  brought forth a powerful tool to study cosmology and astrophysics. Systems where the background source is a quasar can be used to map the dark matter substructure \citep[e.g.][]{MaoSchneider1998,Kochanek+2004,Vegetti+2012,Nierenberg+2014}; to determine the mass \citep[e.g.][]{Morgan+2010} and spin \citep{Reynolds+2014} of black holes; to measure the properties of distant host galaxies \citep[e.g.][]{Kochanek+2001,Claeskens+2006,Peng+2006} and to measure the value of the Hubble constant $H_{0}$. The constraints on cosmological parameters in particular, are comparable in precision to baryonic acoustic oscillation methods \citep[e.g][]{Suyu+2010,Suyu+2013,Bonvin+2016}. In addition to that, the effects of microlensing of the quasar induced by the stars in the lens galaxy can be used to probe the physical properties of quasar accretion disks such as the wavelength dependence of the size of the accretion disk \citep[e.g.][]{Poindexter+2008}. 

Large samples of new quasar lens systems were discovered through dedicated surveys in radio, like the Cosmic Lens All Sky Survey (CLASS; \citealt{Myers+2003,Browne+2003}) that, in combination with the Jodrell Bank VLA Astrometic Survey (JVAS; \citealt{King+1999}), found 21 new lens systems, and in the optical, such as the SDSS Quasar Lens Search (SQLS; \citealt{Oguri+2006,Oguri+2008,Inada+2008,Inada+2010,Inada+2012}), that discovered 49 new lensed systems\footnote{\url{http://www-utap.phys.s.u-tokyo.ac.jp/~sdss/sqls/lens.html}}. The future of the field is currently hindered by the need for more lenses. This can be achieved, according to \citealt{OguriMarshall2010} (hereafter OM10), not by increasing the depth of the searches, but the area. Therefore, surveys such as the Dark Energy Survey (DES - \citealt{DES2005,Abbott+2016}) and, in the future, the Large Synoptic Survey Telescope (LSST - \citealt{Ivezic+2008}) are capable of more than doubling the current quasar lens sample size. 

\begin{figure}
\begin{center}
	\includegraphics[width=1.0\columnwidth]{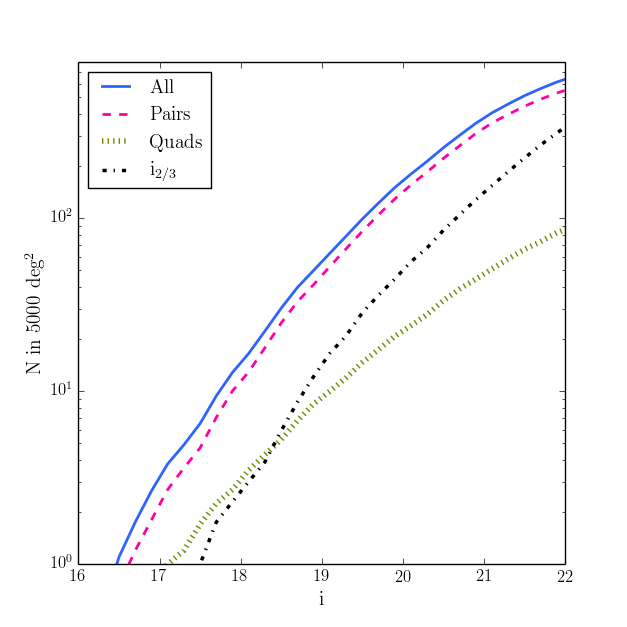}
\end{center}
    \caption{Expected number of quasar lenses in the full DES area as function of the magnitude of the brightest quasar image. The blue solid line shows the overall expected number of systems, while the dashed pink line and the dotted green line show the number of pairs and quads, respectively. The dot-dashed black line shows the expected number of lenses as a function of the magnitude of the second brightest quasar image for pairs and third brightest image for quads.}
    \label{fig:om10predict}
\end{figure}

The chance of a given quasar being lensed was determined by OM10 to be $\sim$10$^{-3.5}$, which is comparable to what was obtained with SQLS, that shows a rate of quasar lensing of $\sim$10$^{-3.3}$. In Fig.~\ref{fig:om10predict} we show the expected number of lenses in the full DES survey area as a function of $i$ band magnitude according to the predictions by OM10. We show the expected numbers for pairs and quads (lenses with four quasar images) according to the magnitude of the brightest image in the system ($i_{1}$). For comparison, we also plotted the expected number of lenses according to the magnitude of the fainter image (for pairs) or the third brightest image (for quads), $i_{2/3}$, which is the limit used by OM10.

With 50 lensed quasar systems with $i_{1}<19.0$ expected in DES, the challenge becomes how to identify them. SQLS started from a spectroscopic sample of quasars. However, the relatively low numbers of confirmed quasars in the Southern Hemisphere sky and the fact that a large spectroscopic survey is not planned for the next few years, requires a method to photometrically select quasars to look for lenses. Traditionally, that selection would rely on the use of the $u$ band to look for UVX objects \citep[e.g.][]{Croom+2001, Richards+2002}. The lack of this band in DES requires the use of the near and mid IR to make the selection. The use of mid-IR has been applied efficiently for flux limited quasar selection \citep[e.g.][]{Stern+2012, Assef+2013} and \cite{DiPompeo+2015} has shown that the use of SDSS+WISE photometry provided results similar to those obtained with SDSS+UV+near-IR data. 

Here, we present results of a search for gravitationally lensed quasars from DES Year 1 observations \citep{Diehl+2014}, obtained between 31 August 2013 and 10 February 2014, combined with JK near infra-red observations from the VISTA Hemisphere Survey (VHS - \citealt{VISTA}; ESO Observing Programme 179.A-2010)  and Wide Infra-red Survey Explorer (WISE - \citealt{WISE}). All magnitudes are quoted on the AB system. The conversions from Vega to AB that have been used for the VISTA data are: $J_{AB} = J_{Vega} + 0.937$ and $Ks_{AB} = Ks_{Vega} + 1.839$. These are taken from the Cambridge Astronomical Survey Unit’s website\footnote{\url{http://casu.ast.cam.ac.uk/surveys-projects/vista/technical/filter-set}}. The conversions for the ALLWISE data are $W1_{AB} = W1_{Vega}+2.699$ and $W2_{AB} = W2_{Vega}+3.339$ which are given in \citet{Jarrett+2011} and in the \textit{ALLWISE} explanatory supplement\footnote{The \textit{ALLWISE} explanatory supplement, \url{http://wise2.ipac.caltech.edu/docs/release/allwise/expsup/sec5\_3e.html}, directs the reader to the \textit{WISE All-Sky} explanatory supplement for the conversions; \url{http://wise2.ipac.caltech.edu/docs/release/allsky/expsup/sec4\_4h.html\#summary}.}. When required, a flat cosmology with $\Omega_{m} = 0.3$ and $H_{0} = 70.0$kms$^{-1}$Mpc$^{-1}$ was used unless otherwise specified.

\section{Candidate selection and modelling}

\subsection{Photometric Data}

Our selection strategy is based on the identification of objects with quasar-like colours that either appear as close pairs or exhibit shape parameters that differ from single point sources. While the motivation for the first criterion is obvious: lensed quasars will appear as multiple images in the sky; the reasoning for the second criterion arises from the fact that, with a median seeing varying between 0.87" and 1.17" across bands, only wider separation lensed quasars in DES will appear as deblended sources. This becomes an even greater problem if the lensing galaxy is bright enough to contaminate the quasar images, posing a complex problem for source segmentation. Even if the individual components of a lensed quasar system are deblended, the pixel assignment to each segmented source is uncertain and the measured catalogue parameters for each quasar image might not manifest as a canonical point source. 

\begin{table}
	\caption{Candidate selection}
	\label{tab:selection}
	\begin{center}
	\begin{tabular}{lr} 
		\hline
		Criterion & Objects remaining \\
		\hline
		DES Y1: $i_{auto}<19.0$ & 7,159,768 \\ 
		Valid $g_{auto}$ & 7,114,967 \\ 
		Match to WISE & 5,156,101 \\ 
		Match to VHS & 4,344,994\\ 
		Valid $JKW1W2$ & 4,171,836\\
		GMM quasars & 19,651 \\
		$i_{psf} - i_{model}>0.1$ & 7,634 \\
		Visual inspection (Eq.~\ref{eq:visinspec}) & 4,778 \\ 
		\hline
	\end{tabular}
	\end{center}
\end{table}

One can define morphology as the difference between $psf$ and $model$ magnitudes in one or more bands: $\mathrm{mag_{psf}} - \mathrm{mag_{model}}$. Effectively, this is the difference between the best fitted local point spread function and the radial light profile of a galaxy. When compared to $\mathrm{mag_{psf}}$, the value of $\mathrm{mag_{model}}$ for extended objects will show an excess since their light profiles will extend beyond what is enclosed by the PSF modelling. However, pixel assignment may wrongly distribute flux from companion sources and hence lead to an overestimation of $\mathrm{mag_{model}}$ and subsequently cause two adjacent point sources to appear non-stellar in the catalogue. Tests performed with lenses simulated according to OM10 mock catalogue of lensed quasars confirm this result. With 4400 lenses with separations varying between $0.5" - 4.0"$ and with $i<21.5$ for the input magnitudes for the source quasar, we find that no system with image separation less than $1.5"$ is segmented and only $23\%$ of pairs are deblended into separate sources. Out of those, only $32\%$ are deblended into point sources. For wider separation lenses, where more massive lensing galaxies can be expected, the contamination by the lensing galaxy on the $model$ magnitude of the quasar images is evident, with only $9\%$ of the systems where $i_{\mathrm{lens}}<20.0$ deblending into point source quasar images. When one looks at systems where $z_{\mathrm{lens}}<1.0$, where the lens galaxy makes a bigger contribution, only $25\%$ of the lenses show point source quasar images.

This calls for a search for objects with quasar colours regardless of image morphology. Ideally, such a technique would also include objects where lensing galaxy flux is present but this is beyond the scope of this paper and will be presented elsewhere  (Ostrovski et al. in prep). Previous quasar selection work such as was done by SDSS focused on selection of point sources prior to the colour selection criterion. Extended sources were selected only if they showed colours that were very different from those of quiescent galaxies \citep{Richards+2002}. Thus lensed quasar surveys such as SQLS will be biased against unresolved candidates. Difference imaging, as suggested by \cite{Kochanek+2006} could also be used with DES multi-epoch observations to identify lensed quasars in those cases where a lensed quasar component exhibits variability.

In our selection we use DES Y1 data from the first annual release (Y1A1) coadded catalogue, which cover an area of approximately 1800 deg$^{2}$ in the Southern Hemisphere sky \citep{Bechtol+2015}. DES images are obtained using the Dark Energy Camera \citep{Flaugher+2015} and reduced through the DES Data Management system \citep{Mohr+2012} using the source extraction code SExtractor \citep{BertinArnouts1996}. Candidates are selected from an input sample composed of the set of all objects brighter than $i_{auto}<19.0$ in the DES Y1 coadded catalogue, that have valid photometry in the $g$ band and are a match to WISE within a 2" search radius. Objects are also matched to VHS, which includes $\sim$1700 deg$^{2}$ of overlap area with DES, to obtain photometry in $J$ and $K$ using a 1.44" search radius. Given the bright limit in our sample, the selection will be performed on a high S/N regime and as such it should lead to good performance. The final $giJKW1W2$ photometric sample on which the gravitationally lensed quasar selection will be performed contains $\sim4.2\times10^{6}$ objects.

The quasar-like colour similarity is calculated in a five-dimensional colour space composed of $g-i$, $i-W1$, $J-K$, $K-W1$ and $W1-W2$. For the DES bands we use $auto$ magnitudes. The $auto$ magnitudes are intended as an estimate of the total flux given by a Kron-like \citep{Kron1980} automatic aperture. While $psf$ magnitudes are optimized to measure the fluxes of stars and $model$ magnitudes are well suited for galaxies, blended quasar lenses won't be well represented by either model. By using a magnitude defined by an elliptical aperture based on the second order moments of the object’s light distribution, at least $90\%$ of the flux should be included and we hope to get well represented colours  independent of object shape. For lensed systems in particular, we want to avoid losing light from one of the components, i.e. having a $psf$ magnitude centered on the lensing galaxy only, and a subsequent misclassification. For the VHS bands we use aperture magnitudes (with an aperture radius $r=\sqrt{2}"$), analogous to the DES SExtractor $auto$ magnitudes. For the lower resolution ($\mathrm{FWHM}=6"$) WISE bands we use the profile-fitting photometry made available in the catalogues. Given the low resolution of the survey, all objects, regardless of morphology appear as point sources.

\subsection{Supervised machine learning}

We apply supervised machine learning to select the candidates. In this work, we use Gaussian Mixture Models (GMM) implemented using astroML \citep{astroml} and scikit-learn \citep{scikit-learn} tools and trained on objects from the Stripe-82 area (S82), a 2.5$^{\circ}$ wide stripe along the Celestial Equator in the Southern Galactic Cap imaged multiple times by SDSS \citep{Abazajian+2009} and other surveys of varying wavelengths. In DES Y1, S82 covers an area of $\sim$167 deg$^{2}$. The S82 sample was selected in a similar way to our input sample, but we required all 5 DES bands to have valid photometry. The resulting sample contains 258,267 objects, 836 of which are spectroscopically confirmed type 1 quasars from the training set used by \cite{Richards+2015} in their classifier. Given the $i<19.0$ magnitude limit we are using and the multi-survey spectroscopic follow-up in S82, we believe we have a complete type 1 quasar sample to train on. Objects in the training set are separated into two classes, quasars and non-quasars, and the GMM method models each class as a set of 10 Gaussians.

Tests to evaluate the method's performance with different combinations of colours were conducted by separating $25\%$ of this training set into a testing sample. Results show that $\sim$88.8$\%$ of the quasars are recovered and the rate of false positives, that is sources that are classified as quasars but are not quasars, is $\sim$27.0$\%$. The classes are assigned based on the highest probability and no minimal thresholds were required. A paper detailing this method is under preparation (Ostrovski et al. in prep).

The GMM returns 19,651 quasar candidates from the input sample, out of which 70 are pairs within 10" of each other. As expected, most of those objects ($\sim$61\%) are point sources. We consider an object to be stellar-like if $i_{\mathrm{psf}} - i_{\mathrm{model}}<0.1$, a more conservative approach than what was done in the SDSS photometric survey \citep{Stoughton+2002}. By removing point sources from our candidate list, we reduce it to 7,634 objects. A quick visual inspection of some of the candidates show that low redshift galaxies are the most obvious contaminants. This group includes starburst galaxies and $z<0.8$ quasars, where, given DES depth, surface brightness limit and resolution, the host galaxy of the AGN becomes prominent. Table~\ref{tab:selection} summarizes the cleaning and selection steps applied to the original DES Y1 sample down to the final visual inspection sample containing 4,478 objects that led to the discovery of J2325-5229. We list the different criteria applied and the remaining number of objects after each step. 

\begin{figure}
   \begin{center}
	\includegraphics[width=1.\columnwidth]{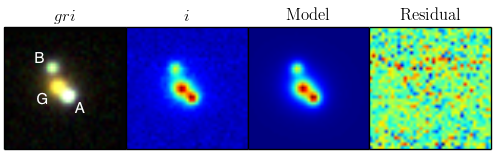}
	\end{center}
    \caption{J2325-5229 as a $g$, $r$, and $i$ DES Y1 colour composite, an $i$ band image, an $i$ band image model and the residuals from subtracting the model from the image. All cutouts are 10.0" in size. North is up and East is left.} 
    \label{fig:residuals}
\end{figure}

\begin{figure*}
\resizebox{1.0\textwidth}{!}{\begin{minipage}{\textwidth}
\begin{center}
	\includegraphics[width=1.0\columnwidth]{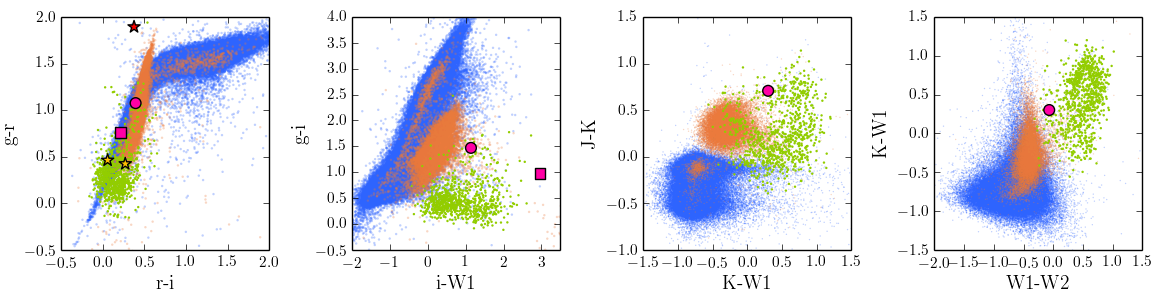}
\end{center}
    \caption{Colour-colour plots showing the location of J2325-5229 (pink circle) in different colour spaces. The pink square shows the location of the north-most blue object. We did not assign W1 flux to each counterpart, so values on the second diagram show, for both cases, the sum of all flux in that band. For comparison, the colour locus of quasars (green), point sources (blue) and extended sources (orange) is populated by the objects present in the GMM training set. The lensing galaxy dominates the colours of the system, making J2325-5229 an outlier of the quasar colour locus. In the left-most plot the stars represent each component of the system after 2D modelling for the photometry (quasar images are yellow and lensing galaxy is red). All magnitudes are in the AB system.}
    \label{fig:allthecolours}
\end{minipage}}
\end{figure*}

Motivated by the morphologies displayed by the simulated OM10 lenses, we selected for visual inspection objects that displayed the following morphological criteria: 
\begin{align}
\label{eq:visinspec}
\left\{
  \begin{array}{lr}
0.2<i_{\mathrm{psf}} - i_{\mathrm{model}}<1.9; \\
\mathrm{ellipticity}<0.3.
\end{array}
\right.
\end{align}
\break\noindent Colour composites made of DES $g$, $r$, and $i$ bands were generated and VDES J2325-5229, shown in Fig.~\ref{fig:residuals}, stood out. The separation between the two blue sources, is $\sim$2.9". Despite clearly having three main components, J2325-5229 is only deblended into a double source in the DES catalogue and appears as a single source in the VHS catalogue. That being the case, the north-most blue object, B, did not match to a VHS counterpart within 1.44" ($0.0004^{\circ}$) and therefore was not in the test sample. This is the reason why J2325-5229 was not flagged as a pair. The candidate selection was solely based on the fact that the south-most component of the system, made of A$+$G, shows quasar colours and an extended morphology and visual inspection was vital. The photometry of this component is dominated by the red central object, as can be seen in the colour-colour plots in Fig.~\ref{fig:allthecolours}. In those plots, the candidate tends to lie closer to the galaxy colour locus, which means that classical colour cuts such as those applied by \cite{Agnello+2015} would not have classified this object as one or multiple quasars. This shows the importance of not applying any sort of pruning to the input sample. The photometry for J2325-5229 is detailed in Table~\ref{tab:phot}. In Fig.~\ref{fig:allthecolours} we also show the photometry for component B, when available.

\begin{table}
	\caption{J2325-5229 photometry$^{\mathrm{[a,b]}}$}
	\label{tab:phot}
	\begin{threeparttable}
	\begin{center}
	\begin{tabular}{ l c c | l c} 
	   \hline
		$g_{\mathrm{A+G; B}}$ & 20.08 $\pm$ 0.01 & 21.41 $\pm$ 0.01 & $J_{\mathrm{A+G+B}}$ & 18.48 $\pm$ 0.02 \\
		$r_{\mathrm{A+G; B}}$ & 19.01 $\pm$ 0.01 & 20.65 $\pm$ 0.01 & $H_{\mathrm{A+G+B}}$ & 18.06 $\pm$ 0.02 \\
		$i_{\mathrm{A+G; B}}$ & 18.61 $\pm$ 0.01 & 20.44 $\pm$ 0.01 & $K_{\mathrm{A+G+B}}$ & 17.78 $\pm$ 0.03 \\
		$z_{\mathrm{A+G; B}}$ & 18.26 $\pm$ 0.01 & 20.09 $\pm$ 0.01 & $W1_{\mathrm{A+G+B}}$ & 17.48 $\pm$ 0.03 \\
		$Y_{\mathrm{A+G; B}}$ & 18.14 $\pm$ 0.04 & 19.95 $\pm$ 0.07 & $W2_{\mathrm{A+G+B}}$ & 17.54 $\pm$ 0.04 \\ 
		\hline
	\end{tabular}	
	\begin{tablenotes}
            \item[a] All quoted magnitudes are AB.
            \item[b] Magnitudes are auto for DES, $\sqrt{2}"$ aperture for VHS and measured with profile-fitting photometry for WISE.
    \end{tablenotes}
    	\end{center}
	\end{threeparttable}
\end{table}

\subsection{2D modelling}
\label{sec:2dmodelling}

\begin{table*}
\resizebox{1.0\textwidth}{!}{\begin{minipage}{\textwidth}
\caption{J2325-5229 modelled positions, photometry and photometric redshift}
	\centering
	\label{tab:allthethings}
	\begin{tabular}{ccccccccccccccc} 
		\hline
		obj & RA & DEC & $\Delta\alpha$" & $\Delta\delta$" & $\delta$R & PA ($^{\circ}$) & $g$ & $r$ & $i$ & $z$ & $Y$ & $K$ & Photo-z \\
		\hline
		G & 23:25:41.20 & -52:29:15.1 & 0.00 & 0.00 & 0.00 & 00.0 & 20.77 & 18.88 & 18.50 & 18.14 & 17.95 & 17.15 & 0.302 \\
		A & 23:25:41.10 & -52:29:15.9 & -0.94 & -0.80 & 1.23 & 229.6 & 20.75 & 20.29 & 20.24 & 20.06 & 20.05 & 19.41 & 2.920 \\
		B & 23:25:41.25  & -52:29:13.5 & 0.49 &  1.72 & 1.79 & 15.9 & 21.41 & 20.99 & 20.72 & 20.57 & 20.53 & 19.71 & 2.960 \\
		A-B & - & - & -1.43 & -2.52 & 2.90 & 29.6  & -0.66  & -0.70  & -0.48  & -0.51  & -0.48  & -0.30 & -   \\
		\hline
	\end{tabular}
	\end{minipage}}
\end{table*}


Once a lensed quasar candidate is selected from visual inspection, the next stage is to perform a 2D modelling of the system to ascertain what model the sources are most consistent with and to obtain photometry for each component. We modelled J2325-5229 in all 5 DES bands and in VHS $K$ band. Image quality made the modelling in the $J$ band unreliable and the resolution from WISE makes it impossible to resolve components with these separations. The positions of each component are listed in Table~\ref{tab:allthethings}. 

The modelling, analogous to what was done by \cite{Auger+2011}, used Python routines to combine two Moffat profile PSFs plus a Sersic profile galaxy convolved with the PSF for the system's image. In Fig.~\ref{fig:residuals} the $i$ band image, the model image and the residual signal obtained from subtracting the two images are displayed. The ellipticity of the lensing galaxy, according to the best model, is $q=0.19\pm0.03$ with a position angle of $70^{\circ}\pm5^{\circ}$. The Sersic index is $3.9\pm0.4$. 

\begin{figure}
	\begin{center}
   \includegraphics[width=0.99\columnwidth]{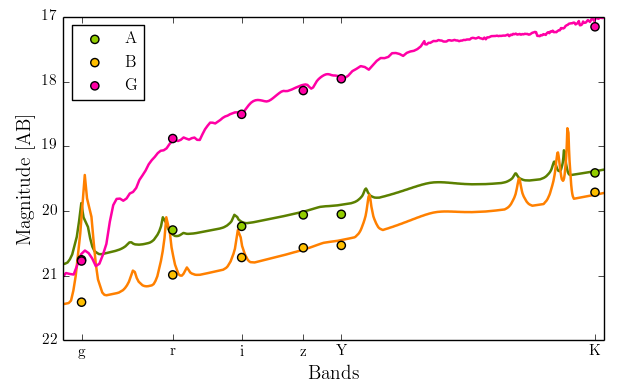}
	\end{center}
    \caption{SEDs of the three components of the J2325-5229 system after modelling the photometry. The continuous lines show the best fit SED template fitted by LePHARE.}
    \label{fig:modelsed}
\end{figure}

For each component, the modelled photometry is listed in Table~\ref{tab:allthethings} and from those values it is possible to plot spectral energy distributions (SED) for each component, which can be seen in Fig.~\ref{fig:modelsed}. As was evident from the colours of the A$+$G blended component, the lensing galaxy is brighter than the quasar images. The difference is of approximately one magnitude in the bluer bands and it can reach almost three magnitudes in the redder bands. 

The modelled colours, displayed on the $gri$ colour-colour diagram in Fig.~\ref{fig:allthecolours}, show that the two blue components are indeed consistent with the quasar locus. The difference between the colours of the quasar-like components can be explained by contamination from the galaxy-like component on the bright counterpart that is projected closer to it. That is supported by the fact that the biggest colour difference is  seen in $r-i$, where the galaxy is brighter, and not in $g-r$. It is also clear, by comparing the modelled photometry of B to that available in the DES catalogue, that despite being segmented, it still contains part of the lensing galaxy flux, particularly in the red bands.

The modelled photometry enables us to calculate photometric redshifts (photo-z) for each component. This was done using the SED fitting code LePHARE (\citealt{Lephare1}; \citealt{Lephare2}). The method compares, through $\chi^{2}$ minimization, the observed magnitudes of a given object with those predicted by the SED which has been convolved with the filter transmission curves. Reddening, interstellar extinction and the opacity of the intergalactic medium are taken into account. 

We used two sets of SED libraries made available with the code: one for galaxy fitting and one for quasars. The first is composed of four observed spectra from \cite{ColemanWuWeedman1980} (linearly extrapolated into ultraviolet and near-IR wavelengths) representing elliptical, irregular, Sbc and Scd galaxies, plus six observed starburst SEDs from the Kinney atlas described in \cite{Calzetti+1994}. The second library is composed of seven synthetic quasar spectra with varying equivalent widths for the Lyman-$\alpha$ and NV lines and both with and without contribution from a blackbody (T=24000K) component. A power law component with spectral indexes of $\alpha=-1.0$ for $597\AA<\lambda<10000\AA$ and $\alpha=-0.7$ for $10000\AA<\lambda<25000\AA$ was used on all quasar spectra. 

For the photometric errors, we estimated 0.1 in each band as an approximation. The best fit photo-z results for each component are listed in Table~\ref{tab:allthethings}. The lensing galaxy was best fit with an elliptical template. Synthetic spectra with different equivalent widths for spectral lines were used to best fit each quasar image, which is not surprising, given that object A is likely to be more heavily contaminated by the lensing galaxy light. The best fit templates are over-plotted in Figure~\ref{fig:modelsed} for each component.

\section{Spectroscopic Observations}

\begin{table}
	\caption{Spectroscopic follow-up of J2325-5229}
	\label{tab:followup}
	\begin{center}
	\begin{tabular}{cccc} 
		\hline
		Telescope & Date & Instrument &  Other\\ 
		\hline
		AAT & 2013 Sep 8/9 & AAOmega & Archival data\\
		NTT & 2015 Oct 7/8 & EFOSC2 & Grism \#13 \\
		\hline
	\end{tabular}
	\end{center}
\end{table}

\subsection{NTT}

Observations were carried out with the ESO Faint Object Spectrograph and Camera 2 (EFOSC2 - \citealt{Buzzoni+1984}) mounted at the f/11 nasmyth focus on the 3.6m NTT on the night of 2015 October 7/8th UT as part of ESO observing program 096-A-041.  The 236lines/mm Grism \#13 blazed at 4400\AA\ was used giving a 5.54\AA/pixel (binned 2x2) and a spectral resolution of 23.0\AA\ for 1'' seeing (FWHM). The full spectral coverage was 3700--9315\AA\ with a delivered spectral resolution (FWHM) of 34.5\AA\ (6 pixels) and 1.5'' seeing during the observations. The spectroscopic slit was aligned with PA = 214$^{\circ}$ so that both candidate quasar components were centred in the slit. No order blocking filter (e.g. GG475) was used since the largest  wavelength range was required for redshift determination and hence there will be to be some second order contamination at wavelengths longer than $2 \times $3700\AA = 7400\AA.

\begin{figure*}
\resizebox{1.0\textwidth}{!}{\begin{minipage}{\textwidth}
\begin{center}
	\includegraphics[width=1.0\columnwidth]{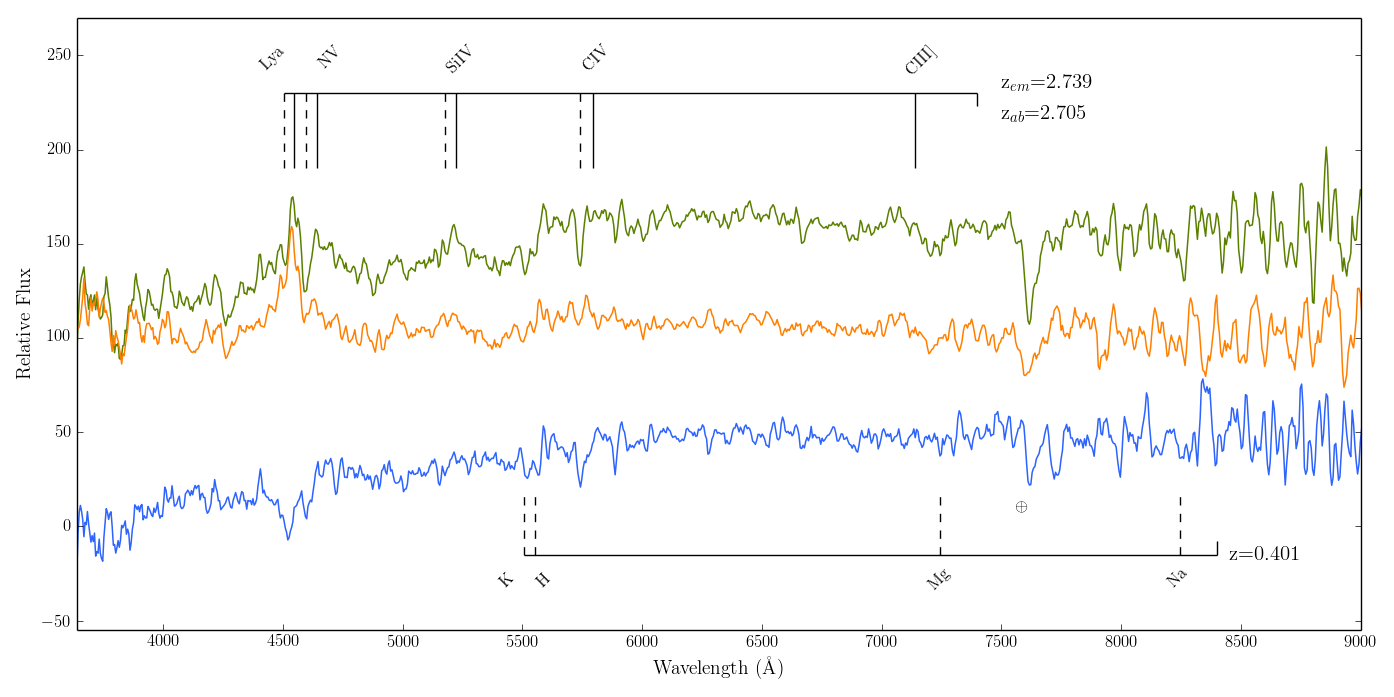}
    \caption{1-D NTT spectra. The top (green) spectrum is the brighter component A and the lower (yellow) spectrum is component B. On top, we have marked the location of the source quasar emission lines at $z_{em}=2.739$ (solid lines) and absorption lines at  $z_{ab}=2.705$ (dashed lines). On the bottom are the lensing galaxy absorption lines at $z_{ab}=0.401$. In blue, we show A-B after scaling the flux of image B by the median flux ratio between the two quasars blueward of $4000\AA$.} 
    \label{fig:ntttspectra}
\end{center}
\end{minipage}}
\end{figure*}

The 2D data were reduced following the guidelines of the PESSTO project \citep{Smartt+2015} using a set of custom python routines with bias frames and flat fields taken during the afternoon prior to the observations. 1D spectra were extracted centred on the two quasar locations with wavelength calibration applied using Arc lamp observations that were were taken immediately after the spectroscopic observations of the science target. The separation of the two quasars is 2.9'' with the galaxy lying between the two quasars and separated 1.2'' from the brighter quasar component A and 1.7'' from component B. Therefore, the galaxy contributes to the spectra of both quasars and is unresolved with
respect to quasar component A in the seeing of 1.5''. The resultant 1D extracted spectra are shown in Figure~\ref{fig:ntttspectra}. 

\subsection{AAT Archival data}

A search for J2325-5229 in other data archives showed it was part of the XXL survey that used the XMM-Newton X-ray telescope to image 50deg$^{2}$ in two fields between 2011 and 2013 \citep{Pierre+2015}. The total X-ray flux for the source is $3.70 \pm 1.10 \times 10^{-14}$erg cm$^{-2}$ s$^{-1}$. The Anglo-Australian Telescope (AAT) was one of the facilities to provide spectroscopic follow-up for the southern XXL field. The follow-up is detailed in \cite{Lidman+2016}, where they describe the use of the two-degree field (2dF) fibre positioner in conjunction with the AAOmega spectrograph with a spectral coverage of 3700--8900$\AA$ and spectral resolution of about 1500 or 4\AA\ at 6000\AA. The data processing is also described.

\begin{figure}
\begin{center}
	\includegraphics[width=1.0\columnwidth]{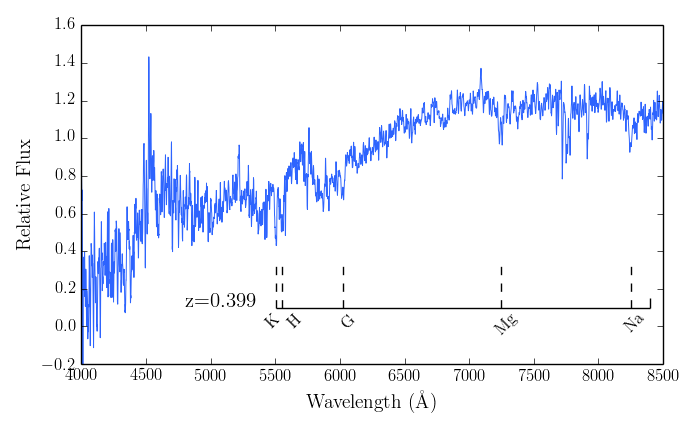}
    \caption{1-D AAT spectrum from archival data. The dashed lines show the location of typical galaxy absorption lines at the given AAT $z=0.399$ redshift.} 
    \label{fig:aatspectra}
\end{center}
\end{figure}

Data for J2325-5229 was obtained between 2013 08 and 09 September and the 1-D spectrum can be seen in Fig~\ref{fig:aatspectra}. The data were processed with 2dfdr\footnote{http://www.aao.gov.au/science/instruments/AAOmega/reduction} through the Australian Dark Energy Survey (OzDES) pipeline \citep{Yuan+2015}. Because the fibre diameter on the sky is $\sim$2" and the separation between the quasar images is $\sim$2.9", most of the flux arises from the lensing galaxy, and the galaxy spectral features are much more prominent in this spectrum than in the NTT spectra. For this reason, the object was not flagged as an AGN in the AAT catalogue. There is, however, clear evidence for quasar features, Lyman-$\alpha$ in particular at $\lambda\sim$4600\AA.

\subsection{Redshift determination}

We measured the wavelength of prominent emission and absorption features in the NTT spectra using the IRAF (v2.16 - \citealt{iraf2}) splot software. The peak or minimum and centroids were measured for each feature. The two measurements generally agreed to within 5\AA\ which is consistent with the spectral resolution (FWHM) of 35\AA\ i.e. 0.2 pixels. Throughout, unless specified, we use rest-frame laboratory wavelengths from \cite{Morton1991} and \cite{TytlerFan1992}.

Both spectra show evidence of a strong emission feature at $\sim$4540\AA\ which we initially identified as redshifted Hydrogen Lyman-$\alpha$ ($\lambda_{rest}= 1215.7\AA$) at an emission line redshift of $z = 2.73$. At this redshift, one expects to see broad NV (1240.1\AA) CIV (1549.1\AA) and CIII] (1908.7\AA) within the observed spectral window at $4631\AA$, 5778\AA\ and 7119\AA\ respectively. Both spectra have potential lines at these wavelengths. In addition, there is evidence of absorption lines blueward of the the NV and CIV permitted lines. No absorption is expected for CIII] since this is a semi-forbidden transition. The observed wavelengths of these features are tabulated in Table~\ref{tab:spectrallinesqso} for each spectrum. We also calculate average emission line and absorption line redshifts excluding the tentative CIII] line. We determine emission line redshifts of 2.739$\pm$0.003 for component A and 2.732 for component B. The absorption line redshifts are 2.705$\pm$0.001 for component A and 2.698$\pm$0.004 for component B, corresponding to a blueshift with respect to the emission line redshifts of 0.034 or 2,700 km/s which is within the range of associated absorption seen in many quasars including mini-BALs. The study of the absorption line velocity and intensity profiles in two sight lines can be used to probe the out-flowing winds of quasars \citep[e.g.][]{Misawa+2013}.

Based on the photometric redshift of 0.302 from Section~\ref{sec:2dmodelling}, we expect the Calcium K and H to have observed wavelengths of 5121.68\AA\ and 5166.99\AA\ respectively. In order to make the galaxy absorption lines more evident in the NTT spectra, we tried to remove the quasar contribution by subtracting the two spectra. But first, we scaled the flux of the fainter image by using the median flux ratio between the two quasars blueward of 4000\AA\, where the galaxy contamination would be negligible. The subtracted spectrum is shown in Figure~\ref{fig:ntttspectra}. There is evidence of a doublet at $\sim$5530\AA\ in both spectra and we identify this as Calcium H and K. We also identify absorption lines consistent with MgI and NaD. The average redshift derived from these lines is 0.401$\pm$0.001. We make
similar measurements from the AAT spectrum and derive a redshift of 0.400$\pm$0.003 as can be seen in Table~\ref{tab:spectrallinesgal}.

Subsequent to our measurements of the AAT spectrum,  \cite{Lidman+2016} published their independent analysis of this data and report a redshift of 0.3996$\pm$0.0003. We take the redshift to be the average of our NTT and AAT analysis at 0.400$\pm$0.002 which is consistent within 0.0004 with the \cite{Lidman+2016} determination. 

\begin{table}
	\caption{Quasar redshift measurements }
	\label{tab:spectrallinesqso}
	\begin{threeparttable}
	\begin{center}
	\begin{tabular}{cccccc} 
		\hline
		& & \multicolumn{2}{ c }{A} & \multicolumn{2}{ c }{B} \\
		Line & $\lambda_{rest} (\AA)$\tnote{c} & $\lambda_{obs} (\AA)$  &  $z$ &  $\lambda_{obs} (\AA)$  & $z$\\ 
		\hline
		Ly$\alpha_{\mathrm{em}}$ & 1215.7 & 4542.8 & 2.737 & 4536.8 & 2.732 \\
		NV$_{\mathrm{ab}}$ & 1240.1 & 4596.5 & 2.706 & 4583.3 & 2.696 \\
		NV$_{\mathrm{em}}$ & 1240.1 & 4639.8 & 2.741 & - & - \\
		CIV$_{\mathrm{ab}}$ & 1549.1 & 5737.8 & 2.704 & 5733.6 & 2.701 \\
		CIII]$_{\mathrm{em}}$ & 1908.7 & 7057.2(?) &  2.697 & 7066.8(?) & 2.703 \\
		\hline
		 &  &  \multicolumn{2}{ c }{$\overline{z}_{\mathrm{em}}=2.739\pm0.003$\tnote{d}} &  \multicolumn{2}{ c }{$\overline{z}_{\mathrm{em}}=2.732$\tnote{d}} \\
		 &  &  \multicolumn{2}{ c }{$\overline{z}_{\mathrm{ab}}=2.705\pm0.001$} &  \multicolumn{2}{ c }{$\overline{z}_{\mathrm{ab}}=2.698\pm0.004$} \\
		\hline
	\end{tabular}
	\begin{tablenotes}
		     \item (?): identification uncertain
            \item[c] Rest frame laboratory wavelengths from Tytler \& Fan 1992
            \item[d] CIII] was not used to calculate average redshifts
    \end{tablenotes}
    	\end{center}
	\end{threeparttable}
\end{table}

\begin{table}
	\caption{Lensing galaxy redshift measurements }
	\label{tab:spectrallinesgal}
	\begin{threeparttable}
	\begin{center}
	\begin{tabular}{cccccc} 
		\hline
		 & & \multicolumn{2}{ c }{NTT} & \multicolumn{2}{ c }{AAT} \\
		Line & $\lambda_{rest} (\AA)$ & $\lambda_{obs} (\AA)$  &  $z$ &  $\lambda_{obs} (\AA)$  & $z$\\ 
		\hline
		Ca K & 3933.7 & 5510.8 & 0.401 & 5506.0 & 0.400 \\
		Ca H & 3968.5 & - & - & 5552.5 & 0.399 \\
		G-band & 4304.0 & - & - & 6018.7 & 0.398 \\
		Mg I & 5175.0 & 7244.5 & 0.400 & 7268.5 &  0.404(?) \\
		Na D & 5892.9 & 8260.4 & 0.402 & 8229.5 & 0.397 \\
		\hline
		 &  &  \multicolumn{2}{ c }{$\overline{z}=0.401\pm0.001$} &  \multicolumn{2}{ c }{$\overline{z}=0.400\pm0.003$} \\
		\hline
	\end{tabular}
	\begin{tablenotes}
            \item (?): identification uncertain
            \item Combined average redshift: $0.400\pm0.002$
    \end{tablenotes}
    	\end{center}
	\end{threeparttable}
\end{table}

\section{Lens Modelling}

Due to the low number of constraints in a double image system, we can only investigate simple lens models, in particular the singular isothermal ellipsoid (SIE).We model the lens system using the public lensing software, GLAFIC (\citealt{Oguri2010ascl}, \citealt{Oguri2010PASJ}). There are eight constraints from observations (positions of the two quasar images and their fluxes, and the position of the lensing galaxy) and eight model parameters (position and flux of the source quasar, position of the lensing galaxy, its mass, ellipticity and position angle). Hence the model has zero degrees of freedom and we expect solutions to converge to $\chi^{2}\sim0$. To estimate uncertainties on the model parameters, we generated 2,000 data sets from the constraints and their errors (as in Table~\ref{tab:allthethings}), including a deviation in the mass density power law of an SIE of 0.1, which covers the observed density distribution in early-type galaxies \citep{Auger+2010}. We use the $i$ band magnitudes as image fluxes and increase the uncertainty on the fluxes by 0.1 mag because of possible contaminations, such as microlensing, dust extinction and intrinsic quasar variability over the time delay \citep[e.g.][]{Hook+1994, Macleod+2012}.

\begin{figure}
\begin{center}
	\includegraphics[width=1.0\columnwidth]{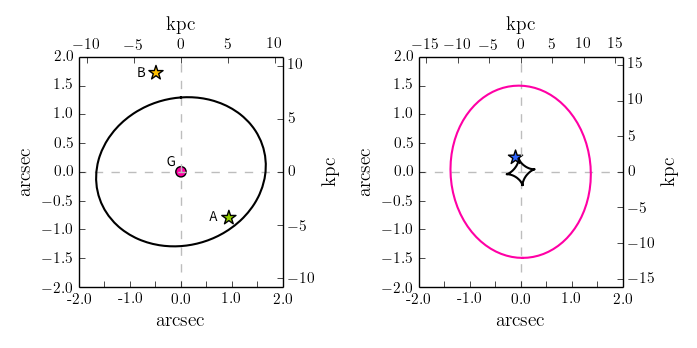}
    \caption{Caustics and critical lines obtained after the mass distribution modelling of J2325. On the left is the image plane, where the black curve indicates the tangential critical line and the pink circle marks out the radial critical curve. Images A and B are the green and yellow stars, respectively. On the right is the source plane, with the radial caustic in pink and the tangential caustic in black. The source quasar is denoted by the blue star.}
    \label{fig:curves}
\end{center}
\end{figure}

In Figure~\ref{fig:curves} we show the diagrams of critical lines and caustics in both the image and the source planes, obtained using GRAVLENS \citep{GRAVLENS}. The results are in agreement with the GLAFIC model. The source quasar is close to the inner caustic but not inside it, which is the reason why the quasar is only doubly lensed. Summarised in Table~\ref{tab:masspars} are the modelling results for Einstein radius ($R_{E}$), ellipticity ($q$), position angle, time delay ($\Delta$t), enclosed mass (M$_{enclosed}$) and magnification ($\mu$). Images A and B are magnified by $-6.5\pm2.2$ and $3.8\pm1.0$ respectively, where the negative sign denotes a parity flip. The time delay was found to be $\Delta$t$=52\pm11$ days (A follows B), given a Hubble constant $H_{0}=68$kms$^{-1}$Mpc$^{-1}$. The discrepancy between position angles measured through the light profile (2D modelling) and mass modelling seen in Table~\ref{tab:masspars} is possibly due to the unknown external shear \citep{Keeton+1998}. 

\begin{table}
	\caption{Parameters of J2325-5229}
	\label{tab:masspars}
	\begin{threeparttable}
	\begin{center}
	\begin{tabular}{ccc} 
		\hline
		 & Lens model & 2D model\\\hline
		$R_{E}$ & $1.47"\pm0.03$ & \\
		q & $0.23\pm 0.06$ & $0.19\pm0.03$ \\
		PA & $97^{\circ}\pm4^{\circ}$ & $70^{\circ}\pm5^{\circ}$  \\
		$\Delta$t\tnote{e} & $52\pm 11$ days  &  \\
		M$_{enclosed}$ & $(4.04\pm 0.15) \times 10^{11}$M$_{\odot}$ & \\
		$\mu$ & $10\pm 3$ &  \\
		\hline
	\end{tabular}
	\begin{tablenotes}
            \item[e] Time delays are positive if A follows B.
    \end{tablenotes}
    	\end{center}
	\end{threeparttable}
\end{table}

\section{Discussion and conclusions}

We have identified a high redshift lensed quasar, VDES J2325-5229, by applying GMM supervised machine learning to select the candidates in a colour space defined by DES+VHS+WISE photometric bands. Since the selection does not depend on the $u$-band, we are capable of selecting candidates at higher redshifts. For comparison, amongst the 49 new lenses found by SQLS, only 6 have source redshifts greater than J2325, 3 of which were serendipitously discovered \citep{Johnston+2003,Pindor+2004,McGreer+2010}. Two more lenses with source redshifts greater than J2325 were discovered in SDSS-III BOSS quasar lens survey \citep{More+2016}, including the highest redshift multiply lensed quasar known, with $z_{s}=4.819$.

Given the geometry of the system, with the presence of an obvious LRG galaxy, and a angular separation between the quasars that converts to a physical distance of $\sim$23.2kpc, it is unlikely that we are looking at distinct binary quasars. The quasar SEDs are considerably similar, with emission line redshifts that differ by $562\pm240$km/s in the quasar rest frame. Given the poor seeing during the NTT spectroscopic observations and the contamination from the lensing galaxy particularly in the SED of quasar component A, that discrepancy is not surprising. Both quasar images present an intrinsic absorption line system detected in CIV and NV, blue shifted by $2739\pm254$km/s in component A and $2744\pm324$km/s in component B with respect to the emission line redshift, further evidencing the SED similarity.

There is evidence that the CIV absorption line is weaker in image B, which would be consistent with different sight lines in the broad-line region of the source quasar as predicted by \cite{Misawa+2014}. This can be seen in Fig~\ref{fig:ntttspectra}, where we have scaled the flux of image B by the median flux ratio between the two quasars blueward of $4000\AA$, where the contamination of the lensing galaxy is most negligible, and subtracted B from A. A strong absorption feature remains at $\lambda\sim5734\AA$, where the CIV absorption is expected to be for $z_{ab}=2.7$. A lesser effect is seen at $\lambda\sim4596\AA$, where one expects the NV absorption to be. This is unsurprising giving the ionization potentials of 64.5eV and 97.9eV for CIV and NV respectively, which means that NV absorbers would be closer to the flux source, and therefore be less likely to be affected by differences in sight lines. Further high-resolution spectroscopy of J2325 would allow for the different sight lines scenario to be confirmed and to constrain the size of the absorber and geometry of the broad-emission line region.

The modelled $i$-band magnitude of the lensing galaxy can be used to estimate the rest frame $R$ band magnitude. At these redshifts, and given our choice of rest and observed frame filters, the $K$-corrections can be neglected. For the LRG in J2325-5229, we calculate $M_{R}=-22.41$, given $z=0.4$ and $m_{i}=18.50$. We can compare this magnitude to that obtained by using the separation of the quasar images and the lensing Faber-Jackson relation with fit parameters described in \cite{Rusin+2003} following what was done in \cite{Jackson+2008}. Assuming $z=0.400\pm0.002$ and $\theta=2.90$,

\begin{equation}
M_{R}=M_{\star R}+2.5\gamma_{E+K}z-1.25\gamma_{FJ}\log{\theta}
\end{equation}

\noindent yields $M_{R}=-23.11\pm0.53$, which is close to the expected value obtained from modelling the data.

The next DES data release will contain data from the full 5000 deg$^{2}$ survey area and will provide a rich sample in which to look for lensed quasars. With such a large area, one can expect to find dozens of bright lenses, including those with quadruple images and the technique introduced in this paper should be able to select all of them as candidates. 

\section*{Acknowledgements}

FO is supported jointly by CAPES (the Science without Borders programme) and the Cambridge Commonwealth Trust.

RGM, CAL, MWA, MB, SLR acknowledge the support of UK Science and Technology Research Council (STFC).

AJC acknowledges the support of a Raymond and Beverly Sackler visiting fellowship at the Institute of Astronomy.

Funding for the DES Projects has been provided by the U.S. Department of Energy, the U.S. National Science Foundation, the Ministry of Science and Education of Spain, 
the Science and Technology Facilities Council of the United Kingdom, the Higher Education Funding Council for England, the National Center for Supercomputing 
Applications at the University of Illinois at Urbana-Champaign, the Kavli Institute of Cosmological Physics at the University of Chicago, 
the Center for Cosmology and Astro-Particle Physics at the Ohio State University,
the Mitchell Institute for Fundamental Physics and Astronomy at Texas A\&M University, Financiadora de Estudos e Projetos, 
Funda{\c c}{\~a}o Carlos Chagas Filho de Amparo {\`a} Pesquisa do Estado do Rio de Janeiro, Conselho Nacional de Desenvolvimento Cient{\'i}fico e Tecnol{\'o}gico and 
the Minist{\'e}rio da Ci{\^e}ncia, Tecnologia e Inova{\c c}{\~a}o, the Deutsche Forschungsgemeinschaft and the Collaborating Institutions in the Dark Energy Survey. 

The Collaborating Institutions are Argonne National Laboratory, the University of California at Santa Cruz, the University of Cambridge, Centro de Investigaciones Energ{\'e}ticas, 
Medioambientales y Tecnol{\'o}gicas-Madrid, the University of Chicago, University College London, the DES-Brazil Consortium, the University of Edinburgh, 
the Eidgen{\"o}ssische Technische Hochschule (ETH) Z{\"u}rich, 
Fermi National Accelerator Laboratory, the University of Illinois at Urbana-Champaign, the Institut de Ci{\`e}ncies de l'Espai (IEEC/CSIC), 
the Institut de F{\'i}sica d'Altes Energies, Lawrence Berkeley National Laboratory, the Ludwig-Maximilians Universit{\"a}t M{\"u}nchen and the associated Excellence Cluster Universe, 
the University of Michigan, the National Optical Astronomy Observatory, the University of Nottingham, The Ohio State University, the University of Pennsylvania, the University of Portsmouth, 
SLAC National Accelerator Laboratory, Stanford University, the University of Sussex, Texas A\&M University, and the OzDES Membership Consortium.

The DES data management system is supported by the National Science Foundation under Grant Number AST-1138766.
The DES participants from Spanish institutions are partially supported by MINECO under grants AYA2012-39559, ESP2013-48274, FPA2013-47986, and Centro de Excelencia Severo Ochoa SEV-2012-0234.
Research leading to these results has received funding from the European Research Council under the European Union’s Seventh Framework Programme (FP7/2007-2013) including ERC grant agreements 
 240672, 291329, and 306478.

The analysis presented here is based on observations obtained as part of the VISTA Hemisphere Survey, ESO Programme, 179.A-2010 (PI: McMahon) and ESO Programme 096.A-0411.

This work was based in part on data acquired through the Australian Astronomical Observatory, under programs A/2013A/018 and A/2013B/001.

This research made use of Astropy \citep{Astropy}, a community-developed core Python package for Astronomy (Astropy Collaboration, 2013).

This research made use of OM10\footnote{https://github.com/drphilmarshall/OM10} mock catalogue of strong gravitational lenses and FO thanks Dr Phil Marshall for support and useful discussions.



\bibliographystyle{mnras/mnras}
\bibliography{references} 

\begin{thebibliography}{}
\makeatletter
\relax
\def\mn@urlcharsother{\let\do\@makeother \do\$\do\&\do\#\do\^\do\_\do\%\do\~}
\def\mn@doi{\begingroup\mn@urlcharsother \@ifnextchar [ {\mn@doi@}
  {\mn@doi@[]}}
\def\mn@doi@[#1]#2{\def\@tempa{#1}\ifx\@tempa\@empty \href
  {http://dx.doi.org/#2} {doi:#2}\else \href {http://dx.doi.org/#2} {#1}\fi
  \endgroup}
\def\mn@eprint#1#2{\mn@eprint@#1:#2::\@nil}
\def\mn@eprint@arXiv#1{\href {http://arxiv.org/abs/#1} {{\tt arXiv:#1}}}
\def\mn@eprint@dblp#1{\href {http://dblp.uni-trier.de/rec/bibtex/#1.xml}
  {dblp:#1}}
\def\mn@eprint@#1:#2:#3:#4\@nil{\def\@tempa {#1}\def\@tempb {#2}\def\@tempc
  {#3}\ifx \@tempc \@empty \let \@tempc \@tempb \let \@tempb \@tempa \fi \ifx
  \@tempb \@empty \def\@tempb {arXiv}\fi \@ifundefined
  {mn@eprint@\@tempb}{\@tempb:\@tempc}{\expandafter \expandafter \csname
  mn@eprint@\@tempb\endcsname \expandafter{\@tempc}}}

\bibitem[\protect\citeauthoryear{{Abazajian} et~al.,}{{Abazajian}
  et~al.}{2009}]{Abazajian+2009}
{Abazajian} K.~N.,  et~al., 2009, \mn@doi [ApJs] {10.1088/0067-0049/182/2/543},
  \href {http://adsabs.harvard.edu/abs/2009ApJS..182..543A} {182, 543}

\bibitem[\protect\citeauthoryear{{Agnello} et~al.,}{{Agnello}
  et~al.}{2015}]{Agnello+2015}
{Agnello} A.,  et~al., 2015, \mn@doi [MNRAS] {10.1093/mnras/stv2171}, \href
  {http://adsabs.harvard.edu/abs/2015MNRAS.454.1260A} {454, 1260}

\bibitem[\protect\citeauthoryear{{Arnouts}, {Cristiani}, {Moscardini},
  {Matarrese}, {Lucchin}, {Fontana}  \& {Giallongo}}{{Arnouts}
  et~al.}{1999}]{Lephare1}
{Arnouts} S.,  {Cristiani} S.,  {Moscardini} L.,  {Matarrese} S.,  {Lucchin}
  F.,  {Fontana} A.,   {Giallongo} E.,  1999, \mn@doi [MNRAS]
  {10.1046/j.1365-8711.1999.02978.x}, \href
  {http://adsabs.harvard.edu/abs/1999MNRAS.310..540A} {310, 540}

\bibitem[\protect\citeauthoryear{{Assef} et~al.,}{{Assef}
  et~al.}{2013}]{Assef+2013}
{Assef} R.~J.,  et~al., 2013, \mn@doi [ApJ] {10.1088/0004-637X/772/1/26}, \href
  {http://adsabs.harvard.edu/abs/2013ApJ...772...26A} {772, 26}

\bibitem[\protect\citeauthoryear{{Astropy Collaboration} et~al.,}{{Astropy
  Collaboration} et~al.}{2013}]{Astropy}
{Astropy Collaboration} et~al., 2013, \mn@doi [\aap]
  {10.1051/0004-6361/201322068}, \href
  {http://adsabs.harvard.edu/abs/2013A%26A...558A..33A} {558, A33}

\bibitem[\protect\citeauthoryear{{Auger}, {Treu}, {Bolton}, {Gavazzi},
  {Koopmans}, {Marshall}, {Moustakas}  \& {Burles}}{{Auger}
  et~al.}{2010}]{Auger+2010}
{Auger} M.~W.,  {Treu} T.,  {Bolton} A.~S.,  {Gavazzi} R.,  {Koopmans}
  L.~V.~E.,  {Marshall} P.~J.,  {Moustakas} L.~A.,   {Burles} S.,  2010,
  \mn@doi [ApJ] {10.1088/0004-637X/724/1/511}, \href
  {http://adsabs.harvard.edu/abs/2010ApJ...724..511A} {724, 511}

\bibitem[\protect\citeauthoryear{{Auger}, {Treu}, {Brewer}  \&
  {Marshall}}{{Auger} et~al.}{2011}]{Auger+2011}
{Auger} M.~W.,  {Treu} T.,  {Brewer} B.~J.,   {Marshall} P.~J.,  2011, \mn@doi
  [MNRAS] {10.1111/j.1745-3933.2010.00980.x}, \href
  {http://adsabs.harvard.edu/abs/2011MNRAS.411L...6A} {411, L6}

\bibitem[\protect\citeauthoryear{{Bechtol} et~al.,}{{Bechtol}
  et~al.}{2015}]{Bechtol+2015}
{Bechtol} K.,  et~al., 2015, \mn@doi [ApJ] {10.1088/0004-637X/807/1/50}, \href
  {http://adsabs.harvard.edu/abs/2015ApJ...807...50B} {807, 50}

\bibitem[\protect\citeauthoryear{{Bertin} \& {Arnouts}}{{Bertin} \&
  {Arnouts}}{1996}]{BertinArnouts1996}
{Bertin} E.,  {Arnouts} S.,  1996, A\&AS, \href
  {http://adsabs.harvard.edu/abs/1996A%26AS..117..393B} {117, 393}

\bibitem[\protect\citeauthoryear{{Bonvin} et~al.,}{{Bonvin}
  et~al.}{2016}]{Bonvin+2016}
{Bonvin} V.,  et~al., 2016, preprint, \href
  {http://adsabs.harvard.edu/abs/2016arXiv160701790B} {} (\mn@eprint {arXiv}
  {1607.01790})

\bibitem[\protect\citeauthoryear{{Browne} et~al.,}{{Browne}
  et~al.}{2003}]{Browne+2003}
{Browne} I.~W.~A.,  et~al., 2003, \mn@doi [MNRAS]
  {10.1046/j.1365-8711.2003.06257.x}, \href
  {http://adsabs.harvard.edu/abs/2003MNRAS.341...13B} {341, 13}

\bibitem[\protect\citeauthoryear{{Buzzoni} et~al.,}{{Buzzoni}
  et~al.}{1984}]{Buzzoni+1984}
{Buzzoni} B.,  et~al., 1984, The Messenger, \href
  {http://esoads.eso.org/abs/1984Msngr..38....9B} {38, 9}

\bibitem[\protect\citeauthoryear{{Calzetti}, {Kinney}  \&
  {Storchi-Bergmann}}{{Calzetti} et~al.}{1994}]{Calzetti+1994}
{Calzetti} D.,  {Kinney} A.~L.,   {Storchi-Bergmann} T.,  1994, \mn@doi [ApJ]
  {10.1086/174346}, \href {http://adsabs.harvard.edu/abs/1994ApJ...429..582C}
  {429, 582}

\bibitem[\protect\citeauthoryear{{Claeskens}, {Sluse}, {Riaud}  \&
  {Surdej}}{{Claeskens} et~al.}{2006}]{Claeskens+2006}
{Claeskens} J.-F.,  {Sluse} D.,  {Riaud} P.,   {Surdej} J.,  2006, \mn@doi
  [\aap] {10.1051/0004-6361:20054352}, \href
  {http://adsabs.harvard.edu/abs/2006A%26A...451..865C} {451, 865}

\bibitem[\protect\citeauthoryear{{Coleman}, {Wu}  \& {Weedman}}{{Coleman}
  et~al.}{1980}]{ColemanWuWeedman1980}
{Coleman} G.~D.,  {Wu} C.-C.,   {Weedman} D.~W.,  1980, \mn@doi [ApJS]
  {10.1086/190674}, \href {http://adsabs.harvard.edu/abs/1980ApJS...43..393C}
  {43, 393}

\bibitem[\protect\citeauthoryear{{Croom}, {Smith}, {Boyle}, {Shanks},
  {Loaring}, {Miller}  \& {Lewis}}{{Croom} et~al.}{2001}]{Croom+2001}
{Croom} S.~M.,  {Smith} R.~J.,  {Boyle} B.~J.,  {Shanks} T.,  {Loaring} N.~S.,
  {Miller} L.,   {Lewis} I.~J.,  2001, \mn@doi [MNRAS]
  {10.1046/j.1365-8711.2001.04474.x}, \href
  {http://adsabs.harvard.edu/abs/2001MNRAS.322L..29C} {322, L29}

\bibitem[\protect\citeauthoryear{{DES Collaboration}}{{DES
  Collaboration}}{2005}]{DES2005}
{DES Collaboration} 2005, ArXiv Astrophysics e-prints, \href
  {http://adsabs.harvard.edu/abs/2005astro.ph.10346T} {}

\bibitem[\protect\citeauthoryear{{DES Collaboration} et~al.,}{{DES
  Collaboration} et~al.}{2016}]{Abbott+2016}
{DES Collaboration} et~al., 2016, \mn@doi [MNRAS] {10.1093/mnras/stw641}, \href
  {http://adsabs.harvard.edu/abs/2016MNRAS.tmp..452D} {}

\bibitem[\protect\citeauthoryear{{DiPompeo}, {Bovy}, {Myers}  \&
  {Lang}}{{DiPompeo} et~al.}{2015}]{DiPompeo+2015}
{DiPompeo} M.~A.,  {Bovy} J.,  {Myers} A.~D.,   {Lang} D.,  2015, \mn@doi
  [MNRAS] {10.1093/mnras/stv1562}, \href
  {http://adsabs.harvard.edu/abs/2015MNRAS.452.3124D} {452, 3124}

\bibitem[\protect\citeauthoryear{{Diehl} et~al.,}{{Diehl}
  et~al.}{2014}]{Diehl+2014}
{Diehl} H.~T.,  et~al., 2014, in Observatory Operations: Strategies, Processes,
  and Systems V. p. 91490V, \mn@doi{10.1117/12.2056982}

\bibitem[\protect\citeauthoryear{{Flaugher} et~al.,}{{Flaugher}
  et~al.}{2015}]{Flaugher+2015}
{Flaugher} B.,  et~al., 2015, \mn@doi [\aj] {10.1088/0004-6256/150/5/150},
  \href {http://adsabs.harvard.edu/abs/2015AJ....150..150F} {150, 150}

\bibitem[\protect\citeauthoryear{{Hook}, {McMahon}, {Boyle}  \& {Irwin}}{{Hook}
  et~al.}{1994}]{Hook+1994}
{Hook} I.~M.,  {McMahon} R.~G.,  {Boyle} B.~J.,   {Irwin} M.~J.,  1994, \mn@doi
  [MNRAS] {10.1093/mnras/268.2.305}, \href
  {http://adsabs.harvard.edu/abs/1994MNRAS.268..305H} {268, 305}

\bibitem[\protect\citeauthoryear{{Ilbert} et~al.,}{{Ilbert}
  et~al.}{2006}]{Lephare2}
{Ilbert} O.,  et~al., 2006, \mn@doi [A\&A] {10.1051/0004-6361:20065138}, \href
  {http://adsabs.harvard.edu/abs/2006A%26A...457..841I} {457, 841}

\bibitem[\protect\citeauthoryear{{Inada} et~al.,}{{Inada}
  et~al.}{2008}]{Inada+2008}
{Inada} N.,  et~al., 2008, \mn@doi [AJ] {10.1088/0004-6256/135/2/496}, \href
  {http://adsabs.harvard.edu/abs/2008AJ....135..496I} {135, 496}

\bibitem[\protect\citeauthoryear{{Inada} et~al.,}{{Inada}
  et~al.}{2010}]{Inada+2010}
{Inada} N.,  et~al., 2010, \mn@doi [AJ] {10.1088/0004-6256/140/2/403}, \href
  {http://adsabs.harvard.edu/abs/2010AJ....140..403I} {140, 403}

\bibitem[\protect\citeauthoryear{{Inada} et~al.,}{{Inada}
  et~al.}{2012}]{Inada+2012}
{Inada} N.,  et~al., 2012, \mn@doi [AJ] {10.1088/0004-6256/143/5/119}, \href
  {http://adsabs.harvard.edu/abs/2012AJ....143..119I} {143, 119}

\bibitem[\protect\citeauthoryear{{Ivezic} et~al.,}{{Ivezic}
  et~al.}{2008}]{Ivezic+2008}
{Ivezic} Z.,  et~al., 2008, \mn@doi [Serbian Astronomical Journal]
  {10.2298/SAJ0876001I}, \href
  {http://adsabs.harvard.edu/abs/2008SerAJ.176....1I} {176, 1}

\bibitem[\protect\citeauthoryear{{Jackson}, {Ofek}  \& {Oguri}}{{Jackson}
  et~al.}{2008}]{Jackson+2008}
{Jackson} N.,  {Ofek} E.~O.,   {Oguri} M.,  2008, \mn@doi [MNRAS]
  {10.1111/j.1365-2966.2008.13268.x}, \href
  {http://adsabs.harvard.edu/abs/2008MNRAS.387..741J} {387, 741}

\bibitem[\protect\citeauthoryear{{Jarrett} et~al.,}{{Jarrett}
  et~al.}{2011}]{Jarrett+2011}
{Jarrett} T.~H.,  et~al., 2011, \mn@doi [ApJ] {10.1088/0004-637X/735/2/112},
  \href {http://adsabs.harvard.edu/abs/2011ApJ...735..112J} {735, 112}

\bibitem[\protect\citeauthoryear{{Johnston} et~al.,}{{Johnston}
  et~al.}{2003}]{Johnston+2003}
{Johnston} D.~E.,  et~al., 2003, \mn@doi [\aj] {10.1086/379001}, \href
  {http://adsabs.harvard.edu/abs/2003AJ....126.2281J} {126, 2281}

\bibitem[\protect\citeauthoryear{{Keeton}}{{Keeton}}{2001}]{GRAVLENS}
{Keeton} C.~R.,  2001, ArXiv Astrophysics e-prints, \href
  {http://adsabs.harvard.edu/abs/2001astro.ph..2340K} {}

\bibitem[\protect\citeauthoryear{{Keeton}, {Kochanek}  \& {Falco}}{{Keeton}
  et~al.}{1998}]{Keeton+1998}
{Keeton} C.~R.,  {Kochanek} C.~S.,   {Falco} E.~E.,  1998, \mn@doi [ApJ]
  {10.1086/306502}, \href {http://adsabs.harvard.edu/abs/1998ApJ...509..561K}
  {509, 561}

\bibitem[\protect\citeauthoryear{{King}, {Browne}, {Marlow}, {Patnaik}  \&
  {Wilkinson}}{{King} et~al.}{1999}]{King+1999}
{King} L.~J.,  {Browne} I.~W.~A.,  {Marlow} D.~R.,  {Patnaik} A.~R.,
  {Wilkinson} P.~N.,  1999, \mn@doi [\mnras]
  {10.1046/j.1365-8711.1999.02328.x}, \href
  {http://adsabs.harvard.edu/abs/1999MNRAS.307..225K} {307, 225}

\bibitem[\protect\citeauthoryear{{Kochanek} \& {Dalal}}{{Kochanek} \&
  {Dalal}}{2004}]{Kochanek+2004}
{Kochanek} C.~S.,  {Dalal} N.,  2004, \mn@doi [ApJ] {10.1086/421436}, \href
  {http://adsabs.harvard.edu/abs/2004ApJ...610...69K} {610, 69}

\bibitem[\protect\citeauthoryear{{Kochanek}, {Keeton}  \& {McLeod}}{{Kochanek}
  et~al.}{2001}]{Kochanek+2001}
{Kochanek} C.~S.,  {Keeton} C.~R.,   {McLeod} B.~A.,  2001, \mn@doi [ApJ]
  {10.1086/318350}, \href {http://adsabs.harvard.edu/abs/2001ApJ...547...50K}
  {547, 50}

\bibitem[\protect\citeauthoryear{{Kochanek}, {Mochejska}, {Morgan}  \&
  {Stanek}}{{Kochanek} et~al.}{2006}]{Kochanek+2006}
{Kochanek} C.~S.,  {Mochejska} B.,  {Morgan} N.~D.,   {Stanek} K.~Z.,  2006,
  \mn@doi [ApJl] {10.1086/500559}, \href
  {http://adsabs.harvard.edu/abs/2006ApJ...637L..73K} {637, L73}

\bibitem[\protect\citeauthoryear{{Kron}}{{Kron}}{1980}]{Kron1980}
{Kron} R.~G.,  1980, \mn@doi [ApJS] {10.1086/190669}, \href
  {http://adsabs.harvard.edu/abs/1980ApJS...43..305K} {43, 305}

\bibitem[\protect\citeauthoryear{{Lidman} et~al.,}{{Lidman}
  et~al.}{2016}]{Lidman+2016}
{Lidman} C.,  et~al., 2016, \mn@doi [PASA] {10.1017/pasa.2015.52}, \href
  {http://adsabs.harvard.edu/abs/2016PASA...33....1L} {33, e001}

\bibitem[\protect\citeauthoryear{{MacLeod} et~al.,}{{MacLeod}
  et~al.}{2012}]{Macleod+2012}
{MacLeod} C.~L.,  et~al., 2012, \mn@doi [ApJ] {10.1088/0004-637X/753/2/106},
  \href {http://adsabs.harvard.edu/abs/2012ApJ...753..106M} {753, 106}

\bibitem[\protect\citeauthoryear{{Mao} \& {Schneider}}{{Mao} \&
  {Schneider}}{1998}]{MaoSchneider1998}
{Mao} S.,  {Schneider} P.,  1998, \mn@doi [\mnras]
  {10.1046/j.1365-8711.1998.01319.x}, \href
  {http://adsabs.harvard.edu/abs/1998MNRAS.295..587M} {295, 587}

\bibitem[\protect\citeauthoryear{{McGreer} et~al.,}{{McGreer}
  et~al.}{2010}]{McGreer+2010}
{McGreer} I.~D.,  et~al., 2010, \mn@doi [\aj] {10.1088/0004-6256/140/2/370},
  \href {http://adsabs.harvard.edu/abs/2010AJ....140..370M} {140, 370}

\bibitem[\protect\citeauthoryear{{McMahon}, {Banerji}, {Gonzalez}, {Koposov},
  {Bejar}, {Lodieu}, {Rebolo}  \& {VHS Collaboration}}{{McMahon}
  et~al.}{2013}]{VISTA}
{McMahon} R.~G.,  {Banerji} M.,  {Gonzalez} E.,  {Koposov} S.~E.,  {Bejar}
  V.~J.,  {Lodieu} N.,  {Rebolo} R.,   {VHS Collaboration} 2013, The Messenger,
  \href {http://adsabs.harvard.edu/abs/2013Msngr.154...35M} {154, 35}

\bibitem[\protect\citeauthoryear{{Misawa}, {Inada}, {Ohsuga}, {Gandhi},
  {Takahashi}  \& {Oguri}}{{Misawa} et~al.}{2013}]{Misawa+2013}
{Misawa} T.,  {Inada} N.,  {Ohsuga} K.,  {Gandhi} P.,  {Takahashi} R.,
  {Oguri} M.,  2013, \mn@doi [\aj] {10.1088/0004-6256/145/2/48}, \href
  {http://adsabs.harvard.edu/abs/2013AJ....145...48M} {145, 48}

\bibitem[\protect\citeauthoryear{{Misawa}, {Inada}, {Oguri}, {Gandhi},
  {Horiuchi}, {Koyamada}  \& {Okamoto}}{{Misawa} et~al.}{2014}]{Misawa+2014}
{Misawa} T.,  {Inada} N.,  {Oguri} M.,  {Gandhi} P.,  {Horiuchi} T.,
  {Koyamada} S.,   {Okamoto} R.,  2014, \mn@doi [ApJl]
  {10.1088/2041-8205/794/2/L20}, \href
  {http://adsabs.harvard.edu/abs/2014ApJ...794L..20M} {794, L20}

\bibitem[\protect\citeauthoryear{{Mohr} et~al.,}{{Mohr}
  et~al.}{2012}]{Mohr+2012}
{Mohr} J.~J.,  et~al., 2012, in Society of Photo-Optical Instrumentation
  Engineers (SPIE) Conference Series.  (\mn@eprint {arXiv} {1207.3189}),
  \mn@doi{10.1117/12.926785}

\bibitem[\protect\citeauthoryear{{More} et~al.,}{{More}
  et~al.}{2016}]{More+2016}
{More} A.,  et~al., 2016, \mn@doi [MNRAS] {10.1093/mnras/stv2813}, \href
  {http://adsabs.harvard.edu/abs/2016MNRAS.456.1595M} {456, 1595}

\bibitem[\protect\citeauthoryear{{Morgan}, {Kochanek}, {Morgan}  \&
  {Falco}}{{Morgan} et~al.}{2010}]{Morgan+2010}
{Morgan} C.~W.,  {Kochanek} C.~S.,  {Morgan} N.~D.,   {Falco} E.~E.,  2010,
  \mn@doi [ApJ] {10.1088/0004-637X/712/2/1129}, \href
  {http://adsabs.harvard.edu/abs/2010ApJ...712.1129M} {712, 1129}

\bibitem[\protect\citeauthoryear{{Morton}}{{Morton}}{1991}]{Morton1991}
{Morton} D.~C.,  1991, \mn@doi [ApJs] {10.1086/191601}, \href
  {http://adsabs.harvard.edu/abs/1991ApJS...77..119M} {77, 119}

\bibitem[\protect\citeauthoryear{{Myers} et~al.,}{{Myers}
  et~al.}{2003}]{Myers+2003}
{Myers} S.~T.,  et~al., 2003, \mn@doi [MNRAS]
  {10.1046/j.1365-8711.2003.06256.x}, \href
  {http://adsabs.harvard.edu/abs/2003MNRAS.341....1M} {341, 1}

\bibitem[\protect\citeauthoryear{{Nierenberg}, {Treu}, {Wright}, {Fassnacht}
  \& {Auger}}{{Nierenberg} et~al.}{2014}]{Nierenberg+2014}
{Nierenberg} A.~M.,  {Treu} T.,  {Wright} S.~A.,  {Fassnacht} C.~D.,   {Auger}
  M.~W.,  2014, \mn@doi [MNRAS] {10.1093/mnras/stu862}, \href
  {http://adsabs.harvard.edu/abs/2014MNRAS.442.2434N} {442, 2434}

\bibitem[\protect\citeauthoryear{{Oguri}}{{Oguri}}{2010a}]{Oguri2010ascl}
{Oguri} M.,  2010a, {glafic: Software Package for Analyzing Gravitational
  Lensing}, Astrophysics Source Code Library (\mn@eprint {ascl} {1010.012})

\bibitem[\protect\citeauthoryear{{Oguri}}{{Oguri}}{2010b}]{Oguri2010PASJ}
{Oguri} M.,  2010b, \mn@doi [PASJ] {10.1093/pasj/62.4.1017}, \href
  {http://adsabs.harvard.edu/abs/2010PASJ...62.1017O} {62, 1017}

\bibitem[\protect\citeauthoryear{{Oguri} \& {Marshall}}{{Oguri} \&
  {Marshall}}{2010}]{OguriMarshall2010}
{Oguri} M.,  {Marshall} P.~J.,  2010, MNRAS, 405, 2579

\bibitem[\protect\citeauthoryear{{Oguri} et~al.,}{{Oguri}
  et~al.}{2006}]{Oguri+2006}
{Oguri} M.,  et~al., 2006, \mn@doi [AJ] {10.1086/506019}, \href
  {http://adsabs.harvard.edu/abs/2006AJ....132..999O} {132, 999}

\bibitem[\protect\citeauthoryear{{Oguri} et~al.,}{{Oguri}
  et~al.}{2008}]{Oguri+2008}
{Oguri} M.,  et~al., 2008, \mn@doi [AJ] {10.1088/0004-6256/135/2/512}, \href
  {http://adsabs.harvard.edu/abs/2008AJ....135..512O} {135, 512}

\bibitem[\protect\citeauthoryear{Pedregosa et~al.,}{Pedregosa
  et~al.}{2011}]{scikit-learn}
Pedregosa F.,  et~al., 2011, Journal of Machine Learning Research, 12, 2825

\bibitem[\protect\citeauthoryear{{Peng}, {Impey}, {Rix}, {Kochanek}, {Keeton},
  {Falco}, {Leh{\'a}r}  \& {McLeod}}{{Peng} et~al.}{2006}]{Peng+2006}
{Peng} C.~Y.,  {Impey} C.~D.,  {Rix} H.-W.,  {Kochanek} C.~S.,  {Keeton} C.~R.,
   {Falco} E.~E.,  {Leh{\'a}r} J.,   {McLeod} B.~A.,  2006, \mn@doi [ApJ]
  {10.1086/506266}, \href {http://adsabs.harvard.edu/abs/2006ApJ...649..616P}
  {649, 616}

\bibitem[\protect\citeauthoryear{{Pierre} et~al.,}{{Pierre}
  et~al.}{2015}]{Pierre+2015}
{Pierre} M.,  et~al., 2015, preprint, \href
  {http://adsabs.harvard.edu/abs/2015arXiv151204317P} {} (\mn@eprint {arXiv}
  {1512.04317})

\bibitem[\protect\citeauthoryear{{Pindor} et~al.,}{{Pindor}
  et~al.}{2004}]{Pindor+2004}
{Pindor} B.,  et~al., 2004, \mn@doi [\aj] {10.1086/381904}, \href
  {http://adsabs.harvard.edu/abs/2004AJ....127.1318P} {127, 1318}

\bibitem[\protect\citeauthoryear{{Poindexter}, {Morgan}  \&
  {Kochanek}}{{Poindexter} et~al.}{2008}]{Poindexter+2008}
{Poindexter} S.,  {Morgan} N.,   {Kochanek} C.~S.,  2008, \mn@doi [ApJ]
  {10.1086/524190}, \href {http://adsabs.harvard.edu/abs/2008ApJ...673...34P}
  {673, 34}

\bibitem[\protect\citeauthoryear{{Reynolds}, {Walton}, {Miller}  \&
  {Reis}}{{Reynolds} et~al.}{2014}]{Reynolds+2014}
{Reynolds} M.~T.,  {Walton} D.~J.,  {Miller} J.~M.,   {Reis} R.~C.,  2014,
  \mn@doi [ApJl] {10.1088/2041-8205/792/1/L19}, \href
  {http://adsabs.harvard.edu/abs/2014ApJ...792L..19R} {792, L19}

\bibitem[\protect\citeauthoryear{{Richards} et~al.,}{{Richards}
  et~al.}{2002}]{Richards+2002}
{Richards} G.~T.,  et~al., 2002, \mn@doi [\aj] {10.1086/340187}, \href
  {http://adsabs.harvard.edu/abs/2002AJ....123.2945R} {123, 2945}

\bibitem[\protect\citeauthoryear{{Richards} et~al.,}{{Richards}
  et~al.}{2015}]{Richards+2015}
{Richards} G.~T.,  et~al., 2015, \mn@doi [ApJS] {10.1088/0067-0049/219/2/39},
  \href {http://adsabs.harvard.edu/abs/2015ApJS..219...39R} {219, 39}

\bibitem[\protect\citeauthoryear{{Rusin} et~al.,}{{Rusin}
  et~al.}{2003}]{Rusin+2003}
{Rusin} D.,  et~al., 2003, \mn@doi [ApJ] {10.1086/346206}, \href
  {http://adsabs.harvard.edu/abs/2003ApJ...587..143R} {587, 143}

\bibitem[\protect\citeauthoryear{{Smartt} et~al.,}{{Smartt}
  et~al.}{2015}]{Smartt+2015}
{Smartt} S.~J.,  et~al., 2015, \mn@doi [\aap] {10.1051/0004-6361/201425237},
  \href {http://adsabs.harvard.edu/abs/2015A%26A...579A..40S} {579, A40}

\bibitem[\protect\citeauthoryear{{Stern} et~al.,}{{Stern}
  et~al.}{2012}]{Stern+2012}
{Stern} D.,  et~al., 2012, \mn@doi [ApJ] {10.1088/0004-637X/753/1/30}, \href
  {http://adsabs.harvard.edu/abs/2012ApJ...753...30S} {753, 30}

\bibitem[\protect\citeauthoryear{{Stoughton} et~al.,}{{Stoughton}
  et~al.}{2002}]{Stoughton+2002}
{Stoughton} C.,  et~al., 2002, \mn@doi [AJ] {10.1086/324741}, \href
  {http://adsabs.harvard.edu/abs/2002AJ....123..485S} {123, 485}

\bibitem[\protect\citeauthoryear{{Suyu}, {Marshall}, {Auger}, {Hilbert},
  {Blandford}, {Koopmans}, {Fassnacht}  \& {Treu}}{{Suyu}
  et~al.}{2010}]{Suyu+2010}
{Suyu} S.~H.,  {Marshall} P.~J.,  {Auger} M.~W.,  {Hilbert} S.,  {Blandford}
  R.~D.,  {Koopmans} L.~V.~E.,  {Fassnacht} C.~D.,   {Treu} T.,  2010, \mn@doi
  [ApJ] {10.1088/0004-637X/711/1/201}, \href
  {http://adsabs.harvard.edu/abs/2010ApJ...711..201S} {711, 201}

\bibitem[\protect\citeauthoryear{{Suyu} et~al.,}{{Suyu}
  et~al.}{2013}]{Suyu+2013}
{Suyu} S.~H.,  et~al., 2013, \mn@doi [ApJ] {10.1088/0004-637X/766/2/70}, \href
  {http://adsabs.harvard.edu/abs/2013ApJ...766...70S} {766, 70}

\bibitem[\protect\citeauthoryear{{Tody}}{{Tody}}{1993}]{iraf2}
{Tody} D.,  1993, in {Hanisch} R.~J.,  {Brissenden} R.~J.~V.,   {Barnes} J.,
  eds,  Astronomical Society of the Pacific Conference Series Vol. 52,
  Astronomical Data Analysis Software and Systems II. p.~173

\bibitem[\protect\citeauthoryear{{Tytler} \& {Fan}}{{Tytler} \&
  {Fan}}{1992}]{TytlerFan1992}
{Tytler} D.,  {Fan} X.-M.,  1992, \mn@doi [ApJs] {10.1086/191642}, \href
  {http://adsabs.harvard.edu/abs/1992ApJS...79....1T} {79, 1}

\bibitem[\protect\citeauthoryear{{Vanderplas}, {Connolly}, {Ivezi{\'c}}  \&
  {Gray}}{{Vanderplas} et~al.}{2012}]{astroml}
{Vanderplas} J.,  {Connolly} A.,  {Ivezi{\'c}} {\v Z}.,   {Gray} A.,  2012, in
  Conference on Intelligent Data Understanding (CIDU). pp 47 --54,
  \mn@doi{10.1109/CIDU.2012.6382200}

\bibitem[\protect\citeauthoryear{{Vegetti}, {Lagattuta}, {McKean}, {Auger},
  {Fassnacht}  \& {Koopmans}}{{Vegetti} et~al.}{2012}]{Vegetti+2012}
{Vegetti} S.,  {Lagattuta} D.~J.,  {McKean} J.~P.,  {Auger} M.~W.,  {Fassnacht}
  C.~D.,   {Koopmans} L.~V.~E.,  2012, \mn@doi [\nat] {10.1038/nature10669},
  \href {http://adsabs.harvard.edu/abs/2012Natur.481..341V} {481, 341}

\bibitem[\protect\citeauthoryear{{Walsh}, {Carswell}  \& {Weymann}}{{Walsh}
  et~al.}{1979}]{Walsh+1979}
{Walsh} D.,  {Carswell} R.~F.,   {Weymann} R.~J.,  1979, \mn@doi [Nature]
  {10.1038/279381a0}, \href {http://adsabs.harvard.edu/abs/1979Natur.279..381W}
  {279, 381}

\bibitem[\protect\citeauthoryear{{Wright} et~al.,}{{Wright}
  et~al.}{2010}]{WISE}
{Wright} E.~L.,  et~al., 2010, \mn@doi [AJ] {10.1088/0004-6256/140/6/1868},
  \href {http://adsabs.harvard.edu/abs/2010AJ....140.1868W} {140, 1868}

\bibitem[\protect\citeauthoryear{{Yuan} et~al.,}{{Yuan}
  et~al.}{2015}]{Yuan+2015}
{Yuan} F.,  et~al., 2015, \mn@doi [MNRAS] {10.1093/mnras/stv1507}, \href
  {http://adsabs.harvard.edu/abs/2015MNRAS.452.3047Y} {452, 3047}

\makeatother
\end{thebibliography}

\section*{Affiliations}
{\small
$^{1}$Institute of Astronomy, University of Cambridge, Madingley Road, Cambridge CB3 0HA, UK\\
$^{2}$Kavli Institute for Cosmology, University of Cambridge, Madingley Road, Cambridge CB3 0HA, UK\\
$^{3}$CAPES Foundation, Ministry of Education of Brazil, Brasília - DF 70040-020, Brazil\\
$^{4}$Department of Astronomy, University of Washington, Seattle, WA 98195, USA\\
$^{5}$DAMTP, Centre for Mathematical Sciences, Wilberforce Road, Cambridge CB3 0WA, UK\\
$^{6}$School of Physics, University of Wollongong, Wollongong, NSW 2522, Australia\\
$^{7}$Australian Astronomical Observatory, North Ryde, NSW, Australia\\
$^{8}$Fermi National Accelerator Laboratory, P. O. Box 500, Batavia, IL 60510, USA\\
$^{9}$CNRS, UMR 7095, Institut d'Astrophysique de Paris, F-75014, Paris, France\\
$^{10}$Department of Physics \& Astronomy, University College London, Gower Street, London, WC1E 6BT, UK\\
$^{11}$Sorbonne Universit\'es, UPMC Univ Paris 06, UMR 7095, Institut d'Astrophysique de Paris, F-75014, Paris, France\\
$^{12}$Laborat\'orio Interinstitucional de e-Astronomia - LIneA, Rua Gal. Jos\'e Cristino 77, Rio de Janeiro, RJ - 20921-400, Brazil\\
$^{13}$Observat\'orio Nacional, Rua Gal. Jos\'e Cristino 77, Rio de Janeiro, RJ - 20921-400, Brazil\\
$^{14}$Department of Astronomy, University of Illinois, 1002 W. Green Street, Urbana, IL 61801, USA\\
$^{15}$National Center for Supercomputing Applications, 1205 West Clark St., Urbana, IL 61801, USA\\
$^{16}$Institut de Ci\`encies de l'Espai, IEEC-CSIC, Campus UAB, Carrer de Can Magrans, s/n,  08193 Bellaterra, Barcelona, Spain\\
$^{17}$Institut de F\'{\i}sica d'Altes Energies (IFAE), The Barcelona Institute of Science and Technology, Campus UAB, 08193 Bellaterra (Barcelona) Spain\\
$^{18}$Kavli Institute for Particle Astrophysics \& Cosmology, P. O. Box 2450, Stanford University, Stanford, CA 94305, USA\\
$^{19}$Excellence Cluster Universe, Boltzmannstr.\ 2, 85748 Garching, Germany\\
$^{20}$Faculty of Physics, Ludwig-Maximilians-Universit\""at, Scheinerstr. 1, 81679 Munich, Germany\\
$^{21}$Department of Astronomy, University of Michigan, Ann Arbor, MI 48109, USA\\
$^{22}$Department of Physics, University of Michigan, Ann Arbor, MI 48109, USA\\
$^{23}$Kavli Institute for Cosmological Physics, University of Chicago, Chicago, IL 60637, USA\\
$^{24}$Department of Astronomy, University of California, Berkeley,  501 Campbell Hall, Berkeley, CA 94720, USA\\
$^{25}$Lawrence Berkeley National Laboratory, 1 Cyclotron Road, Berkeley, CA 94720, USA\\
$^{26}$SLAC National Accelerator Laboratory, Menlo Park, CA 94025, USA\\
$^{27}$Center for Cosmology and Astro-Particle Physics, The Ohio State University, Columbus, OH 43210, USA\\
$^{28}$Department of Physics, The Ohio State University, Columbus, OH 43210, USA\\
$^{29}$Cerro Tololo Inter-American Observatory, National Optical Astronomy Observatory, Casilla 603, La Serena, Chile\\
$^{30}$Australian Astronomical Observatory, North Ryde, NSW 2113, Australia\\
$^{31}$Departamento de F\'{\i}sica Matem\'atica,  Instituto de F\'{\i}sica, Universidade de S\~ao Paulo,  CP 66318, CEP 05314-970, S\~ao Paulo, SP,  Brazil\\
$^{32}$George P. and Cynthia Woods Mitchell Institute for Fundamental Physics and Astronomy, and Department of Physics and Astronomy, Texas A\&M University, College Station, TX 77843,  USA\\
$^{33}$Department of Astronomy, The Ohio State University, Columbus, OH 43210, USA\\
$^{34}$Department of Astrophysical Sciences, Princeton University, Peyton Hall, Princeton, NJ 08544, USA\\
$^{35}$Instituci\'o Catalana de Recerca i Estudis Avan\c{c}ats, E-08010 Barcelona, Spain\\
$^{36}$Jet Propulsion Laboratory, California Institute of Technology, 4800 Oak Grove Dr., Pasadena, CA 91109, USA\\
$^{37}$Department of Physics and Astronomy, Pevensey Building, University of Sussex, Brighton, BN1 9QH, UK\\
$^{38}$Centro de Investigaciones Energ\'eticas, Medioambientales y Tecnol\'ogicas (CIEMAT), Madrid, Spain\\
$^{39}$Instituto de F\'\i sica, UFRGS, Caixa Postal 15051, Porto Alegre, RS - 91501-970, Brazil\\
$^{40}$ICTP South American Institute for Fundamental Research\\ Instituto de F\'{\i}sica Te\'orica, Universidade Estadual Paulista, S\~ao Paulo, Brazil\\
$^{41}$Computer Science and Mathematics Division, Oak Ridge National Laboratory, Oak Ridge, TN 37831\\
$^{42}$Institute of Cosmology \& Gravitation, University of Portsmouth, Portsmouth, PO1 3FX, UK\\
}




\bsp	
\label{lastpage}
\end{document}